\documentstyle[12pt]{article}
\def\strut{\rule[-.5cm]{0cm}{1cm}}

\begin{document}

\title{A Predictive Inflationary Scenario Without The Gauge Singlet}

\author{{\bf G. Lazarides}\\Physics Division, School of Technology\\
University of Thessaloniki\\Thessaloniki, GREECE\\\\{\bf Q. Shafi}\\
Bartol Research Institute\\University of Delaware\\ Newark, Delaware
19716\thanks{Supported in part by DOE Grant No. DE-FG02-91ER40626 and
NATO Grant No. CRG-910573}}

\date{ }
\maketitle

\begin{abstract}
We propose a new realization of the chaotic inflationary scenario in
which the scalar field responsible for inflation also spontaneously
breaks the underlying gauge symmetry at a superheavy scale $\sim
10^{15} - 10^{17}\; GeV$. A possible framework is provided by
the superstring inspired gauge models, in which case several predictions are
essentially model independent. The spectral index for the scalar
perturbations $n \simeq 0.92 - 0.88$, while the ratio of the tensor
to the scalar quadrupole anisotropy is $(\Delta T/T)^2_T/(\Delta T/T)^2_S
\approx
0.4 - 0.7$. On smaller angular scales, $\delta T/T (1^\circ ) \approx (9 - 16)
\times 10^{-6}$ and $\delta T/T (2.1^\circ ) \approx (6 - 10) \times
10^{-6}$. Implications for magnetic monopoles and cosmic strings as
well as the gauge hierarchy problem are pointed out.
\end{abstract}
\newpage

%\dspace
The simplest realizations of the new or the chaotic inflationary
scenario$^{(1)}$ invoke a weakly coupled scalar field which typically
is a singlet under the full gauge symmetry of the model.$^{(2,3)}$
Gauge singlets arise naturally within the framework of higher
dimensional cosmologies and inflationary scenarios based on these
Kaluza-Klein type ideas have been discussed quite some time ago.$^{(4)}$
For almost a decade, however, the most promising approach for
unifying grand unification with gravity has been based on superstring
theories. A variety of related ideas have been explored in the
literature. One of the most elegant is also the earliest,$^{(5)}$
based on the compactification of six of the ten dimensions of the
heterotic $E_8 \times E_8$ superstring theory$^{(6)}$ on a Calabi-Yau
(C-Y) manifold. Four dimensional gauge models obtained from the C-Y approach
are in
many ways more satisfying than the standard supersymmetric grand
unified theories (SUSY GUTS). For instance, the number of chiral
families is related to the Euler character of the underlying C-Y
manifold.$^{(5)}$ Discrete symmetries such as matter parity arise
automatically, and are needed to adequately suppress proton decay,
ensure the existence of a pair of light higgs doublets, etc.

The main purpose of this paper is to point out an intriguing possibility,
that the (chaotic) inflationary
scenario perhaps could be implemented in superstring models without
invoking a gauge singlet field!$^{(7)}$ The idea is that inflation could be
driven by the very same field which also is responsible for breaking
the underlying gauge symmetry at some superheavy scale $\sim 10^{15}
- 10^{17}\; GeV$. The presence of supersymmetry, broken at a scale
$\sim 10^{3\pm 1}\; GeV$, is essential to ensuring the appearance of a
`sufficiently flat' potential, needed both for phenomenology as well
as for inflation. Moreover, the dimensionless coupling which enters in the
determination of $\delta \rho/\rho$ is related to non-perturbative
(instanton) effects of the underlying string theory, thereby
providing (at least) another rationale as to why it happens to be
much smaller than unity. The form of the inflationary
potential in this class of models is narrowly constrained, thereby
allowing for several essentially model independent
predictions.

In the C-Y approach to the heterotic $E_8 \times E_8$ superstring
compactification,$^{(5)}$ the ten-dimensional spacetime splits into $M_4
\times K$, where $M_4$ denotes the Minkowski spacetime and $K$ is a
six-dimensional C-Y manifold. By embedding the spin connection of $K$
in the first $E_8$, one obtains a four-dimensional gauge model which
possesses an $E_6$ gauge symmetry and $N = 1$ supersymmetry. The C-Y
space $K$ is usually constructed as $K = K_0/G$, where $G$ is a
freely acting discrete group on the simply connected C-Y space $K_0$.
The number of chiral fermion families of $E_6$ is given by $\mid
\chi (K_0)\mid /2n(G)$, where $\chi (K_0)$ is the Euler character of $K_0$
and $n(G)$ is the order of $G$. Non-trivial Wilson loops on $K$ will
break $E_6$ to some subgroup $H$ of rank five or six.  Note that in
the compactification schemes under consideration the gauge coupling
typically gets related to the vacuum expectation value of some
dilaton field. Precisely how this occurs, especially in the context
of the early universe is an issue beyond the scope of this paper.

Consider, for definiteness, the case where $H$ is the maximal subgroup $SU(3)_c
\times
SU(3)_L \times SU(3)_R$ of $E_6$. The left-handed lepton, quark and
antiquark fields from the ${\bf 27}$ of $E_6$ transform
under $H$ as

\begin{equation}
\begin{array}{lclcl}
\ell & = & (1,\bar{3},3) & = & \left( \begin{array}{lll}
H^u & H^d & L\\
E^c & \nu^c & N \end{array} \right)\strut\\
q & = & (3,3,1) & = & \left( \begin{array}{c}
u\\ d\\ g \end{array} \right)\strut\\
Q & = & (\bar{3},1,\bar{3}) & = & \left( \begin{array}{c}
u^c\\ d^c\\ g^c \end{array} \right) \end{array}
\end{equation}

\noindent
After the flux breaking, in addition to the three (chiral)
{\bf 27}'s, there should survive at least two massless $\ell,\;
\bar{\ell}$ pairs, to provide the necessary Higgs fields
$N,\bar{N}$ and $\nu^c, \bar{\nu}^c$ for the symmetry
breaking of $(SU(3))^3$ to the minimal supersymmetric standard model.

A variety of observational constraints such as the suppression of
proton decay, $\sin^2\theta_W \simeq 0.23$, etc. require that the
breaking of $SU(3)^3$ takes place at some superheavy scale. In order
to generate this scale, the potential for the fields that acquire vevs
of this order must possess $D$ and $F$ flat directions.$^{(8,9,10)}$ It turns
out
that the $\ell^3$ and $\bar{\ell}^3$ terms (from
$({\bf 27})^3$ and $({\bf \overline{27}})^3)$ in the superpotential are
automatically $F$-flat in the $N,\bar{N}, \nu^c, \bar{\nu}^c$ directions.
To ensure the vanishing of the $D$-terms, pairs of $\ell, \bar{\ell}$ must
acquire vevs along the $N,\bar{N}$ and $\nu^c,\bar{\nu}^c$
directions such that

\begin{equation}
<\sum_i {\bf 27}_i^\dagger T^a {\bf 27}_i> = <\sum_i \overline{{\bf
27}}_i^\dagger T^a \overline{{\bf 27}}_i>\end{equation}

\noindent
This ensures the cancellation of quartic contributions to the
potential from the $D$-terms.

The quartic (leading non-renormalizable) contribution to the superpotential
$W$ takes the generic form

\begin{equation}
\frac{\lambda}{M_P} ({\bf 27}\; \overline{{\bf 27}})^2
\end{equation}

\noindent
where $M_P \simeq 1.2 \times 10^{19} GeV$ is the Planck scale, and $\lambda$ is
a
dimensionless parameter. It has been pointed out$^{(11)}$ that there should be
$F$-flat
directions along which the non-renormalizable contribution in (3)is generated
only through the non-perturbative
world-sheet instanton effects. The coefficient $\lambda$ is then
proportional to $\exp (-c/g^2) (c>0$ and $g$ denotes the world sheet
coupling) and could reasonably be expected to be much smaller than
unity. We will be more precise about the value of $\lambda$ when we discuss the
inflationary aspects of the model.

Next we make the standard assumption that the symmetry breaking of $SU(3)^3$
to the MSSM has a radiative origin. This requires that the
superpotential contains cubic couplings of $N,\bar{N} (\nu^c, \bar{\nu}^c)$
that are sufficiently strong. The coupling $gg^c N (\bar{g} \bar{g}^c
\bar{N})$ is one such
example. The presence of these couplings will ensure that the loop
corrections will drive the mass squared term for        the $N,\bar{N}
(\nu^c, \bar{\nu}^c)$ pair, arising from supersymmetry
breaking,  to the negative values needed for the        spontaneous
symmetry breaking. To simplify, we henceforth base our
discussion on a pair of scalar fields. [For instance, $\phi, \bar{\phi}$
could be the pair $N,\bar{N}$ which breaks $SU(3)^3$ to $SU(3)_c
\times SU(2)_L \times SU(2)_R \times U(1)_{B-L}$]. In the D-flat
direction the scalar potential $V(\phi )$ has the form

\begin{equation}
V(\phi) \approx - M_S^2 \mid \phi\mid^2 + \frac{\lambda^2}{3}
\frac{\mid\phi\mid^6}{M_P^2}\end{equation}

\noindent
where $M_S\; (\sim 10^{3\pm 1}\; GeV)$ denotes the supersymmetry breaking scale
and we assume
that the coefficients of the higher order terms are small enough
to make them negligible during the relevant last stages of the inflationary
phase. [It
remains to be seen whether this important assumption can be realized in
realistic
`string derived' models.] Minimization of $V(\phi)$ gives

\begin{equation}
\mid<\phi>\mid = \mid<\bar{\phi}>\mid \equiv M \simeq \lambda^{-\frac{1}{2}}
(M_PM_S)^\frac{1}{2}
GeV
\end{equation}

For values of $\phi$ larger than
$M$, the $\phi^6$ term
in (4) dominates. Provided that $\lambda$ is sufficiently small, this
kind of potential will yield the chaotic
inflationary scenario.$^{(1)}$ The inflationary phase takes place for $\phi
\gg M_P$ (with the constraint that $\lambda^2\mid\phi\mid^6/3 M_P^2
\stackrel{_<}{_\sim} M_P^4)$, and ends when $\phi$ becomes of order
$0.5 M_P$ (for the $\phi^6$ potential). The field $\phi$ then rolls
down to the minimum at $\phi = M$ and performs damped oscillations of
frequency $\sim M_S$.

An estimate of the order of magnitude of $\lambda$ is obtained by
considering the contribution of the scalar metric perturbation to the
microwave background quadrupole anisotropy (scalar Sachs-Wolfe effect) and
comparing it with the
recent COBE measurement.$^{(12)}$ One has$^{(1)}$ (the subscript $S$
denotes the scalar contribution):

\begin{equation}
\left. \left( \frac{\Delta T}{T}\right)_S^2 \simeq \frac{32\pi}{45}
\frac{V^3}{V^{\prime 2} M_P^6} \right|_{k\sim H}
\end{equation}

\noindent
where the right hand side is to be evaluated when the scale
$k^{-1}$, corresponding to the present horizon size, crossed inside
the horizon during inflation.
Equation (6) can be re-written as

\begin{equation}
\left|\left( \frac{\Delta T}{T}\right)_S \right| \simeq 0.023 \lambda
N_H^2
\end{equation}

\noindent
where $N_H = \frac{2\pi}{3} (\phi/M_P)^2_{k\sim H}$ denotes the
corresponding number of
e-foldings. Taking $N_H$ on the order of 50, [this is somewhat
smaller than the usually quoted value of 60 due to the lower damping
rate of the oscillating $\phi$ field], and $\Delta T/T \approx
6 \times 10^{-6}$ from COBE, we estimate the fundamental quantity
$\lambda$ to be of order $10^{-7}$.

Inserting this value of $\lambda$ in (5), we see that the vev $M \simeq
10^{14.5}.\linebreak (M_S/10^3\; GeV)^\frac{1}{2}\; GeV$. In order to estimate
the
decay width of $\phi$ one needs to know the relevant couplings.
Clearly, since the decay products have masses $\stackrel{_<}{_\sim}
M_S\; (\approx M_\phi)$, these couplings all arise from the
non-renormalizable terms (with suppressed couplings) in the superpotential.
Some typical ones are $H\bar{H} \frac{\phi^2}{M_P}, \nu^c \bar{\nu}^c
\frac{\phi^2}{M_P}$, etc. The first one produces higgsinos as decay
products, while the second coupling gives rise to `right handed'
neutrinos. The decay width $\Gamma$ of $\phi$ is (roughly) estimated to be

\begin{equation}\begin{array}{lcl}
\Gamma & \sim & {\cal O} (10^{-1}) (M_S^3/M^2)\strut\\
& \sim & {\cal O} (10^{-21}) (M_S/10^3\; GeV)^2 GeV\end{array}
\end{equation}

\noindent
The oscillations of $\phi$ are damped out when the Hubble time $t$
becomes $\sim \Gamma^{-1}$, and the universe `reheats' to a
temperature

\begin{equation}
T_r \sim (\Gamma M_P)^{\frac{1}{2}} \sim 10^{-1} (M_S/10^3\; GeV) GeV
\end{equation}

An inflationary scenario is certainly incomplete without an
explanation of the origin of the observed baryon asymmetry in the
universe.  This is a particularly pressing issue in the present case.
The reheat temperature is quite low $(\stackrel{_<}{_\sim} {\rm few} GeV)$,
 so that some of the more interesting (from the inflationary viewpoint)
scenarios, such as baryons from leptons$^{(13)}$ or electroweak
baryogenesis,$^{(14)}$ are not applicable.  Actually, the problem has been
discussed in some
detail in an earlier work.$^{(15)}$  Here, for completeness, we present only
the essential
idea, keeping details to a minimum.  The baryon
asymmetry is given by the formula

\begin{equation}
n_b/s \sim \frac{T_r}{M_\phi} \frac{\Gamma_{\Delta
B\neq0}}{\Gamma}
\end{equation}

\noindent
where $\Gamma (\Gamma_{\Delta B \neq 0})$ denotes the total (baryon
number violating) decay width of $\phi$. Consider the superpotential
couplings $gg^c\phi,\; g^cu^cd^c$ and
$gd^c\nu^c$. The coefficient in front of the first coupling is
assumed to be of order $M_g/<\phi>$, while the remaining two
couplings carry coefficients of order unity. The decay width for
the baryon number violating process $\phi
\rightarrow u^c d^c d^c \nu^c$ is then given by

\begin{equation}\begin{array}{lcl}
\Gamma_{\Delta B\neq0} & \sim & \frac{1}{16\pi} \left(
\frac{1}{8\pi^2}\right)^3 \left( \frac{M_g}{<\phi>}\right)^2
\frac{M_\phi^5}{M_g^4}\cdot\;\; {\rm (no.\; of\; channels)}\strut\\
& \sim & 10^{-4} \frac{M_\phi^5}{<\phi>^2M_g^2}\end{array}
\end{equation}

\noindent
Here $M_g$ denotes the mass of the $g$ boson and in estimating the
number of channels we include the sum over color and flavors. The
baryon asymmetry is estimated to be

\begin{equation}
n_b/s \sim {\cal O} (10^{-1}) \left(
\frac{1}{8\pi^2}\right)^3 \frac{T_rM_\phi}{M_g^2}
\end{equation}

\noindent
A number of comments are in order:

\begin{enumerate}
\item The $g$ boson mass should be $\sim 10^5 - 10^6
GeV$ in order to generate $n_b/s \sim 10^{-10}-10^{-11}$. The
scenario actually requires (a minimum of) two species of $g$'s.
Precise details would be model dependent.

\item The presence of such relatively `light' $g$'s would impose
constraints on the model arising from proton decay, $n-\bar{n}$
oscillations, etc.

\item Since the reheat is so low, 2-2 scatterings cannot wipe out the
asymmetry generated above. This is certainly a plus for the model.
\end{enumerate}

A second scenario for implementing  baryogenesis at low
$(\sim GeV$ - few $MeV)$ temperature has previously been discussed
in ref (16). It needs the presence in the
superpotential of the baryon number violating operator $u^cd^cd^c$
which presumably is a mild requirement. This scenario also seems to
fit well with the present inflationary framework.

We now turn to the important issue of topological defects. Depending
on the model, magnetic
monopoles and/or cosmic strings will arise through the Kibble
mechanism$^{(17)}$ at the end of inflation. For
instance, the breaking of $SU(3)^3$ produces magnetic monopoles.
Remarkably, however, there is no monopole problem.$^{(18)}$
Two crucial differences from ordinary GUTS are i) the
higgs correlation length $\xi$ is of order $M_S^{-1}$ and not $M^{-1}$, and ii)
$\phi$ dominates for quite some time after the production of the
topological defects.

Monopoles are produced via the Kibble mechanism when the $\phi$-field
oscillations over the barrier at $\phi = 0$ with height $M_S^2 M^2$ come
to a halt. Their initial number
density is

\begin{equation}
n_M \sim \frac{p}{\frac{4\pi}{3} \xi^3} \sim 10^{-2} M_S^3
\end{equation}

\noindent
where $p \sim 10^{-1}$ is a geometric factor. Consequently, the
initial monopole energy density is given by

\begin{equation}
\frac{\rho_M}{\rho_\phi} \sim 10^{-2} \frac{M_S^3 m_M}{M_S^2
<\phi>^2} \sim 10^{-2} \frac{M_Sm_M}{<\phi>^2}
\end{equation}

\noindent
where $m_M$ denotes the monopole mass. The ratio in (14) remains
constant until radiation takes over at $T_r$. Assuming that no
further entropy is generated, one finds that $r \equiv n_M/s \sim
10^{-2} \frac{M_ST_r}{<\phi>^2} \sim 10^{-29}$ for $M_S \sim TeV,\;
<\phi> \sim 10^{15}\; GeV$.

We therefore conclude that if the inflationary scenario is
implemented within an $(SU(3))^3$ model, one expects to see magnetic
monopoles (carrying three quanta of Dirac magnetic charge) at or
close to the Parker bound!

Depending on the model, cosmic strings can be produced (analogous to
the magnetic monopoles) at the end of the inflationary epoch. Their
thickness is of order $M_S^{-1}$, while their mass per unit length is
of order $M^2$. One needs $M \sim 10^{16}\; GeV$ for strings to play
a significant role in large scale structure formation. This can be
achieved by
choosing $\lambda$ to be somewhat smaller than $10^{-7}$, in which
case the main source of primordial density fluctuations would be due
to cosmic strings.

Before proceeding to a discussion of some model
independent predictions of this
inflationary scenario, we wish to go back to the superpotential in
this class of models. With some clever symmetries it is possible, in
principle, to eliminate the
lowest order non-renormalizable term in (3). In this case the leading
non-renormalizable term in the superpotential is ${\cal O}
(\frac{1}{M_P^3}) \lambda^\prime ({\bf 27}\; {\bf \overline{27}})^3$. We then
expect the
effective potential $V(\phi)$ to have the form (assuming that the higher
order terms can be ignored; see remarks immediately preceding eq.
(5)):

\begin{equation}
V(\phi) \approx - M^2_S \mid\phi\mid^2 + \frac{\lambda^{\prime 2}}{5}
\frac{\mid\phi\mid^{10}}{M_P^6}
\end{equation}

\noindent
Minimization then yields the vev to be

\begin{equation}
\mid<\phi>\mid = \mid<\bar{\phi}>\mid \equiv M^\prime \simeq
\lambda^{\prime-\frac{1}{4}} (M_S^2
M_P^6)^\frac{1}{8} GeV
\end{equation}

The quantity $\left( \frac{\Delta T}{T}\right)_S$ in this case is
proportional to $N_k^3$. Proceeding as in the previous case, one finds
that the dimensionless parameter $\lambda^\prime \approx 0.2 \times
10^{-8}$. Substitution in (16) yields

\begin{equation}
\mid<\phi>\mid = \mid<\bar{\phi}>\mid = M^\prime \simeq 10^{17} \left(
M_S/10^3\; GeV \right)^\frac{1}{4}\; GeV
\end{equation}

The scale in (17) is somewhat larger than the typical SUSY GUT scale
of $10^{16}\; GeV$, although this need not be an issue. However, the `reheat'
temperature is in the MeV range
at best, and so it should be clear that in order to have the standard
nucleosynthesis senario the leading non-renormalizable terms in the
superpotential should
not be of dimension higher than seven. The effective potential
during the inflationary phase therefore will be assumed to be proportional
either
to $\phi^6$ or $\phi^{10}$. Note that in the latter case the
topological defects become less interesting. The monopole number
density will be extremely small (see eq. (14)), while the cosmic
strings are excessively massive.

In chaotic inflation with a $\phi^\gamma$ scalar potential, the
amplitude of the density perturbation on a given scale $k^{-1}$ as
it crosses inside the horizon is proportional to
$N_k^{\frac{\gamma + 2}{4}} (\approx N_H^{\frac{\gamma + 2}{4}}
(k^{-1}(Mpc)/10^4)^\frac{\gamma + 2}{4 N_H}$). Taking $N_H \approx
50$, this implies that the spectral index $n \approx 0.92 (0.88)$ for
$\gamma = 6 (10)$. Recall that $n = 1$ corresponds to the
Harrison-Zeldovich case.

Employing the well known relation for the gravitational wave
contribution to the quadrupole anisotropy$^{(19)}$

\begin{equation}
(\Delta T/T)^2_T \simeq 0.6 \frac{V}{M_P^4}
\end{equation}

\noindent
we find

\begin{equation}\begin{array}{lcl}
r & \equiv & \frac{(\Delta T/T)^2_T}{(\Delta T/T)^2_S} \approx
\frac{3.4 \gamma}{N_H}\strut\\
& \simeq & 0.4 (0.7)\; {\rm for}\;  \gamma = 6 (10)\end{array}
\end{equation}

Knowing $n$ and $r$ we can estimate the bias factor $b ( \equiv
\sigma_8^{-1}$, where $\sigma_8$ is the rms mass fluctuation on the
scale $8h^{-1}$ Mpc) for a cold dark matter scenario using the approximate
relation$^{(20)}$

\begin{equation}
b_{CDM} \approx 100^{(1-n)/2} \sqrt{1 + r} \approx 1.4 - 1.7
\end{equation}

For a mixed (cold $+ 20\%$ hot) dark matter scenario,$^{(21,22)}$ the
bias factor turns out to be $b_{MDM} \approx 1.5 b_{CDM}$.

Our final topic concerns the anisotropy predictions on angular scales
of $1^\circ$ and $2.1^\circ$. We will follow ref. (23), taking into
account the following. Firstly, it has been noted$^{(24)}$ that
the COBE DMR gives $Q_{rms-PS} \approx 14 \mu K \pm 27\%$, which
corresponds to a reduction of the published COBE numbers by $\approx
15\%$. Secondly, the power spectrum here has less power on smaller scales
and in addition, the tensor contribution to the quadrupole anisotropy is
not negligible.
Taking all this into account we find that, unless reionization was
important, $\delta T/T (1^\circ) \approx (9 - 16) \times 10^{-6}$ and
$\delta T/T(2.1^\circ) \approx (6 - 10) \times 10^{-6}$.

To conclude, the proposal outlined above for implementing inflation could,
also, in principle, be considered within the framework of
ordinary supersymmetric GUTS. One would have to ensure, through
suitable symmetries, that the inflationary potential is
consistent with all of the phenomenological constraints. The superstring
framework (Calabi-Yau, orbifolds, four dimensional constructions,...)
appears, however, to provide a more natural framework. The value of
the dimensionless coupling $\lambda$ (or $\lambda^\prime)$ [and also
presumably of other couplings associated with the leading
non-renormalizable terms in the superpotential], is determined to be
$\sim 10^{-7} - 10^{-8}$, which is precisely what one needs to ensure
the existence of a pair of `light' $(\sim M_S)$ higgs doublets
(assuming, of course, that the doublets acquire their mass only through
the quartic non-renormalizable couplings). Consequently, the doublets
should be protected from acquiring large masses through cubic
couplings in order to resolve the gauge hierarchy
problem. Finally, we have concentrated in this work
on outlining the scenario and describing some model independent
predictions. It would be extremely interesting to find realistic
examples of models in which the coupling $\lambda$ turns out to be of
the right order of magnitude.
\vspace{.2in}

\noindent
\underline{\bf Acknowledgements}: We thank Andrei Linde and Bob Schaefer for
important discussions.
\newpage

\section*{References}

\begin{enumerate}

\item For an authoritative discussion and references to the original
literature see A. Linde, Particle Physics and Inflationary Cosmology,
Harwood Academic Publishers (1990). Also see E.W. Kolb and M.S.
Turner, The Early Universe, Addison-Wesley (1990).

\item Q. Shafi and A. Vilenkin, Phys. Rev. Lett., \underline{52}
(1984) 691.

\item S.Y. Pi, Phys. Rev. Lett., \underline{52} (1984) 1725.

\item Q. Shafi and C. Wetterich, Phys. Lett., \underline{129B} (1983)
387.

\item P. Candelas, G. Horowitz, A. Strominger and E. Witten, Nucl.
Phys. \underline{B258} (1985) 46;\\
E. Witten, Nucl. Phys. \underline{B258} (1985) 75.

\item D. Gross, J. Harvey, E. Martinec and R. Rohm, Phys. Rev. Lett.,
\underline{56} (1985) 502.

\item For an attempt at `new' inflation without the singlet using
preonic theories see M. Cvetic, T. H\"{u}bsch, J.C. Pati and H.
Stremnitzer, Phys. Rev., \underline{D40} (1989) 1311.

\item M. Dine, V. Kaplunovsky, M. Mangano, C.R. Nappi and N. Seiberg,
Nucl. Phys., \underline{B259} (1985) 549.

\item B.R. Greene, K.H. Kirklin, P.J. Miron and G.G. Ross, Nucl.
Phys., \underline{B292} (1987) 606;\\
S. Kalara and R.N. Mohapatra, Phys. Rev., \underline{D36} (1987)
3674;\\
R. Arnowitt and P. Nath, Phys. Rev., \underline{D40} (1989) 191.

\item G. Lazarides, P.K. Mohapatra, C. Panagiotakopoulos and Q.
Shafi, Nucl. Phys., \underline{B323} (1989) 614;\\
B. Ananthanarayan, Q. Shafi and G. Lazarides, Nucl. Phys.,
\underline{B339} (1990) 67.

\item E. Witten, Nucl. Phys., \underline{B268} (1986) 79;\\
M. Dine and C. Lee, Phys. Lett., 203B (1988) 371;\\
M. Cvetic, Phys. Rev., D37 (1988) 2366.

\item G. Smoot, et al., Ap. J. Lett., \underline{396} (1992) L1.

\item G. Lazarides and Q. Shafi, Phys. Lett., \underline{B258} (1991)
305;\\
M. Fukugita and T. Yanagida, Phys. Lett., \underline{B174} (1986) 45.

\item V. Kuzmin, V. Rubakov and M. Shaposhnikov, Phys. Lett.,
\underline{B155} (1985) 36.

\item G. Lazarides, C. Panagiotakopoulos and Q. Shafi, Nucl. Phys.,
\underline{B307} (1988) 937.

\item S. Dimopoulos and L.J. Hall, Phys. Lett., \underline{B196}
(1987) 135.

\item T.W.B. Kibble, J. Phys., \underline{A9} (1976) 1387.

\item G. Lazarides, C. Panagiotakopoulos and Q. Shafi, Phys. Rev.
Lett., \underline{58} (1987) 1707.

\item A.A. Starobinsky, Sov. Astron. Lett., \underline{11} (1985)
113;\\
L. Abbott and M. Wise, Nucl. Phys., \underline{B244} (1984) 541;\\
V. Rubakov, M. Sazhin and A. Veryaskin, Phys. Lett., \underline{B115}
(1982) 189.

\item R.L. Davis, H.M. Hodges, G.F. Smoot, P.J. Steinhardt and M.S.
Turner, Phys. Rev. Lett., \underline{69} (1992) 1856;
Also A. Liddle and D. Lyth, Univ. of Sussex preprint 1992.

\item Q. Shafi and F. Stecker, Phys. Rev. Lett., \underline{53}
(1984) 1292;\\
R. Schaefer, Q. Shafi and F. Stecker, Ap. J., \underline{367} (1989)
575;\\
J.A. Holtzman, Ap. J. Suppl., \underline{71} (1989) 1.

\item A. Van Dalen and R. Schaefer, Ap.J., \underline{398} (1992) 33;\\
M. David, F. Summers and M. Schlegel, Nature, \underline{359} (1992)
393;\\
A. Klypin, J. Holtzman, J. Primack and E. Regos, Univ. of Santa Cruz,
preprint, Nov. 1992;\\
A. Taylor and M. Rowan-Robinson, Nature, \underline{359} (1992) 335.

\item R. Schaefer and Q. Shafi, Phys. Rev. D, \underline{47} (1993)
1333;\\
ibid, Nature, \underline{359} (1992) 359.

\item J.R. Bond, Talk presented at the CMBR Workshop, Berkeley, Dec.
11-12 (1992).
\end{enumerate}

\end{document}